# DEEP NEURAL NETWORK (DNN) FOR WATER/FAT SEPARATION: SUPERVISED TRAINING, UNSUPERVISED TRAINING, AND NO TRAINING

Ramin Jafari, MS[1,2], Pascal Spincemaille, PhD[2], Jinwei Zhang, BS[1,2], Thanh D. Nguyen, PhD[2], Xianfu Luo, MD[2,3], Junghun Cho, PhD[1,2], Daniel Margolis, MD[2], Martin R. Prince, MD, PhD, FACR[2], Yi Wang, PhD[1,2]

Meinig School of Biomedical Engineering, Cornell University, Ithaca, NY [1]

Department of Radiology, Weill Cornell Medicine, New York, NY [2]

Department of Radiology, Northern Jiangsu People's Hospital, Yangzhou, China [3]

**Corresponding author:**

E-mail: rj259@cornell.edu

**Grant Support**: This research was supported in part by NIH grants R01CA181566, R01NS090464, R01DK116126, and R01NS095562.

**Running title**: Unsupervised deep learning for water/fat separation

**ABSTRACT**

**Purpose:** To use a deep neural network (DNN) for solving the optimization problem of water/fat separation and to compare supervised and unsupervised training.

**Methods:** The current T2*-IDEAL algorithm for solving fat/water separation is dependent on initialization. Recently, deep neural networks (DNN) have been proposed to solve fat/water separation without the need for suitable initialization. However, this approach requires supervised training of DNN (STD) using the reference fat/water separation images. Here we propose two novel DNN water/fat separation methods 1) unsupervised training of DNN (UTD) using the physical forward problem as the cost function during training, and 2) no-training of DNN (NTD) using physical cost and backpropagation to directly reconstruct a single dataset. The STD, UTD and NTD methods were compared with the reference T2*-IDEAL.

**Results:** All DNN methods generated consistent water/fat separation results that agreed well with T2*-IDEAL under proper initialization.

**Conclusion:** The water/fat separation problem can be solved using unsupervised deep neural networks.

**Keywords;** Deep Learning, Unsupervised, Label Free, Water/Fat Separation,

**INTRODUCTION**

R2* corrected water/fat separation estimating fat, water and inhomogeneous field from gradient-recalled echo (GRE) is a necessary step in quantitative susceptibility mapping to remove the associated chemical shift contribution to the field (1-5). Several algorithms, including hierarchical multiresolution separation, multi-step adaptive fitting, and T2*-IDEAL, have been proposed to decompose the water/fat separation problem into linear (water and fat) and nonlinear (field and R2*) subproblems and solve these problems iteratively (6-8). Water/fat separation is a nonlinear nonconvex problem that requires a suitable initial guess to converge to a global minimum. Multiple solutions including 2D and 3D graph-cuts and in-phase echo-based acquisition have been proposed to generate an initial guess (3,9-11). The performances of these methods is dependent on the assumptions inherent in these methods, including single species dominant voxels, field smoothness, or fixed echo spacing to generate a suitable initial guess and avoid water/fat swapping (12).

Recently, deep neural networks (DNN) have been used to perform water/fat separation using conventional supervised training of DNN with reference data (STD) (13,14). This STD water/fat separation method does not require an initial guess with the potential use of fewer echoes to shorten the scan time, or improve SNR with the same scan time, and lessen dependency on acquisition parameters compared to current standard approaches (13,14). However, the training of STD requires reference fat-water reconstructions (labels), which can be challenging to calculate as discussed above (15).

In this work, we investigate an unsupervised training of DNN (UTD) method that uses the physical forward problem in T2*-IDEAL as the cost function during conventional training without a need for reference fat-water reconstructions (no labels). We further investigate no-training DNN (NTD) method using a cost function similar to that in unsupervised to reconstruct water-fat images directly from a single dataset. We compare the results of the STD, UTD, and NTD methods with current T2*-IDEAL method.

## MATERIALS AND METHODS

### *Water/Fat Separation T2\*-IDEAL*

Water/fat separation and field estimation is a nonconvex problem of modeling multi-echo complex GRE signal ($S$) in terms of water content ($W$), fat content ($F$), field ($f$) and $R_2^*$ decay per voxel (8):

$$(W, F, f, R_2^*) = \underset{W,F,f,R_2^*}{argmin} \sum_{j=1}^{N} \left\| S_j - e^{-R_2^* t_j} e^{-i2\pi f t_j} (W + F e^{-i2\pi \nu_F t_j}) \right\|_2^2, \quad [1]$$

where $N$ refers to the number of echoes, $S_j$ is the GRE signal $S$ at echo time $t_j$ with $j = 1, \ldots, N$, and $\nu_F$ is the fat chemical shift in a single-peak model. T2*-IDEAL solves Eq.1 by decomposing into linear ($W, F$) and nonlinear ($f, R_2^*$) subproblems. With an initial guess for $f$ and $R_2^*$, the linear subproblem for $W$ and $F$ can be solved. With updated $W, F$, the nonlinear subproblem for $f$ and $R_2^*$ is linearized through first order Taylor expansion and solved using Gauss-Newton optimization (8). These steps are repeated until convergence is achieved. In this study, initial guesses for the field $f$ and $R_2^*$ decay were generated using in-phase echoes (3). The subsequent solutions to Eq. 1 resulted in the reference reconstructions $\Psi_{REF}(S) = \{W, F, f, R_2^*\}$.

### *Supervised Training DNN (STD) Water/Fat Separation*

In this work, we adapted the approaches in recent works (13,14) and making $W, F, f$, and $R_2^*$ the target output of the network. A U-net type network $\Psi(S^i; \theta)$ with network weights $\theta$ is trained on $M$ training pairs $\{S^i, \Psi_{REF}(S^i)\}$, where $S^i$ and $\Psi_{REF}(S^i) = \{W, F, f, R_2^*\}$, are the input complex gradient echo signal and the corresponding reference T2*-IDEAL reconstruction (reference), respectively. The cost function is given by:

$$E_{STD} = \frac{1}{2} \sum_{i=1}^{M} \left\| \Psi_{REF}(S^i) - \Psi(S^i; \theta) \right\|_2^2. \quad [2]$$

### *Unsupervised Training DNN (UTD) Water/Fat Separation*

In the proposed method, termed unsupervised, we seek to use deep learning framework to calculate $W, F, f$, and $R_2^*$ without access to reference reconstructions (labels). This is done by using the physical forward problem in Eq. 1 as the cost function during training. This cost function is given by:

$$E_{UTD} = \frac{1}{2}\sum_{i=1}^{M}\sum_{j=1}^{N}\left\|S_j^i - \tilde{S}_j\left(\Psi(S^i;\theta)\right)\right\|_2^2, \quad [3]$$

with $\tilde{S}_j(\Psi) = e^{-R_2^* t_j}e^{-i2\pi f t_j}(W + Fe^{-i2\pi v_F t_j})$ for $\Psi = \{W, F, f, R_2^*\}$.

*No-Training DNN (NTD) Water/Fat Separation*

Recently, a single test data is used to update DNN weights in deep image prior (16) and fidelity imposed network edit (17). This inspires the idea that DNN weights $\theta^*$ initialized randomly may be updated on a single gradient echo data set $S$ to minimize the cost function in Eq.3 in a single data set $S$.

$$E_{NTD} = \frac{1}{2}\sum_{j=1}^{N}\left\|S_j - \tilde{S}_j\left(\Psi(S_j;\theta)\right)\right\|_2^2 \quad [4]$$

The resulting network weights are specific to the data $S$, and the resulting output $\Psi(S;\theta^*)$ can be taken as the water/fat separation reconstruction of $S$. In contrast to the above STD and UTD that involve conventional training data, no training is required here, the cost function is the same as that in the unsupervised training. Therefore, this method is referred to as no-training DNN (NTD).

*Network Architecture*

The network $\Psi(S^i;\theta)$ was a fully 2D convolutional neural network with encoding and decoding paths (Figure 1). The encoding path included repeated blocks (n=5) each consists of convolution (2×2), activation function (Sigmoid), batch normalization and max pooling (2×2). The decoding path with repeated blocks (n=5) has similar architecture except max pooling is replaced with deconvolution and upsampling along with concatenation of corresponding feature maps with the encoding path [4]. The last bock consists of convolution with linear activation function. The input to the network consisted of $2N$ channels (the magnitude and phase of the gradient echo signal for each echo). The output of the network consisted of 6 channels (the magnitude and phase of the water and fat images, plus the field map and R2*). Figure 1 shows the network architecture, representative input and output images for a test set in UTD method, along with outputs of intermediate layers. These show how learned features at different levels transform the input data into the final output images. Note that yellow arrows indicate concatenation of encoder and decoder layer outputs with the same feature maps. Training was performed using the ADAM optimizer (18).

*Data Acquisition*

Data was acquired in healthy volunteers (n=12) and patients (n=19), including thalassemia major (n=11), polycystic kidney disease (n=7), and suspected iron overload (n=1). The study was approved by the Institutional Review Board and written informed consent was obtained from each participant.

Two 1.5T GE scanners (Signa HDxt, GE Healthcare, Waukesha, WI) with 8-channel cardiac coil were used to acquire data. The healthy subjects were imaged on both scanners using identical protocols. Patients were scanned on one scanner, selected at random, using the same protocol. This protocol contained a multi-echo 3D GRE sequence with the following imaging parameters: number of echoes = 6, unipolar readout gradients, flip angle = 5°, $\Delta TE$ = 2.3 msec, $TR$ = 14.6 msec, acquired voxel size = 1.56×1.56×5 mm$^3$, $BW$ = 488 Hz/pixel, reconstruction matrix = 256×256×(32-36), ASSET acceleration factor = 1.25, and acquisition time of 20-27 sec.

*Experiments and Quantitative Analysis*

Multiple experiments were performed to assess the performance of DNN in water/fat separation and how supervised (STD), unsupervised (UTD), and no-training (NTD), compare against T2*-IDEAL reference method. The network architecture for $\Psi(S^i; \theta)$ was identical between the three DNN methods. Parameters in STD and UTD include, number of epochs 2000, batch size 2, learning rate 0.001 for STD and 0.0001 for UTD. The learning rates were experimentally found to produce optimal results. In NTD, parameters include, number of epochs 10000, batch size 2, and learning rate 0.0001.

The supervised and unsupervised DNN were trained on the combined data set of healthy subjects (2 scans each) and patients. This data of n=43 scans was split into testing (256x256x61x6) and training (256x256x1522x6) with 80% of the latter used to training and the rest for validation. The weights corresponding to the lowest validation loss during training was selected as the optimal weights to be used during testing. Training time in each epoch was ~60 seconds. The testing data comprised of two datasets, one healthy subject and one patient with iron overload. In the test data, ROIs were drawn on the proton density fat fraction map $PDFF = \frac{|F|}{|F|+|W|}$, the field and R2* in several regions including liver, adipose and visceral fat, aorta, spleen, kidney, vertebrae. The ROI measurements for each DNN method were compared with those measured on the reference T2*-IDEAL maps using correlation analysis. Reference generation with T2*-IDEAL was performed on CPU (Inter i7-5820k, 3.3 GHz, 64 GB,) using MATLAB (MathWorks, Natick, MA) and all DNN trainings were performed on GPU (NVIDIA, Titan XP GP102, 1405 MHz, 12 GB) using Keras/TensorFlow.

**RESULTS**

Figure 2 compares the network output results in the healthy test subject for STD (Figure 2b), UTD (Figure 2c), NTD (Figure 2d) with the reference T2*-IDEAL reconstruction (Figure 2a). The correlation plots with T2*-IDEAL for STD, UTD, and NTD in PDFF (Figure 2e), field (Figure 2f) and R2* (Figure 2g) show excellent agreement with slopes close to 1 and $R^2 \geq 0.98$. Supplemtary Figure S1 shows the corresponding result when, instead of magnitude and phase, the real and imaginary components of the gradient echo signal are use as input for the network. While in this case water and fat images agree well with reference, field and R2* shows poor qualitative (Figure S1b-d) and quantitative (Figure S1e-g) agreement with the reference. This suggests while both formats (real/imaginary or magnitude/phase) have identical information content, spatial and temporal distribution of information differs from one format to another which makes the learning task easier in the latter case.

Figure 3 compares the network output results in the moderate iron-overload test patient for STD (Figure 3b), UTD (Figure 3c), NTD (Figure 3d) with the reference T2*-IDEAL reconstruction (Figure 3a). Very good qualitative agreement in both contrast and details is observed among these outputs. There were only marginal deviations in the field of STD in the anterior part of the abdomen near the kidney (yellow arrow in Figure 3b), and marginal differences in R2* maps among the reference and STD, in comparison with UTD (yellow arrow in Figure 3c) and NTD (yellow arrow in Figure 3d). Correlation analysis comparing PDFF (Figure 3e), field (Figure 3f) and R2* (Figure 3g) with the reference images (Figure 3a) show excellent agreement with slopes close to 1 and $R^2 \geq 0.96$.

Figure 4 shows the normalized training and validations losses for the STD and UTD methods. While validation loss for the STD method is initially (epoch < 500) lower than the training loss it becomes larger at later epochs while the opposite trend is observed for UTD. The total training time for both methods was similar (~33 hrs each).

Figure 5 shows a comparison of NTD with T2*-IDEAL. The T2*-IDEAL cost function per iteration is shown in the same graph as the NTD reconstruction cost per epoch. In the first row, T2*-IDEAL results of field (Figure 5a), water (Figure 5b), R2* (Figure 5c) and fat (Figure 5d) shows partial failure in several regions (yellow arrows) when field and R2* were initialized with zeros. In the second row, the corresponding images show the result of T2*-IDEAL when using a proper initialization (the field and R2* obtained from the in-phase echoes in this case). The third row shows the proposed NTD results. The generated maps when using 10000 epochs are close the successful T2*-IDEAL result (2$^{nd}$ row) without a need for an initial guess. The corresponding T2*-IDEAL costs and NTD reconstruction loss are shown in Figure 5e. Without initialization, T2*-IDEAL requires more iterations (10000) compared to T2*-IDEAL with initialization (100). The proposed NTD method requires 10000 epochs for convergence. Computation time for each iteration and each epoch was similar (~2 seconds).

**DISCUSSION AND CONCLUSION**

Our results demonstrate the feasibility of an unsupervised deep neural network (DNN) framework with conventional training and without training to solve the water/fat separation problem. The proposed unsupervised training DNN (UTD) method does not require reference images as in supervised training DNN (STD), allowing the use of DNNs for training data that are unlabeled but for which physical model is known. The no-training DNN (NTD) method further allows using DNN reconstruction of a single data set (subject) for which a physical model is known.

For the nonlinear nonconvex water/fat separation problem, the reference T2*-IDEAL method used traditional gradient descent optimization and is dependent on the initial guess. This problem may be alleviated using deep learning, as long as the labeled training data is sufficiently large to capture test data characteristics. Labeled data may be difficult to obtain, in as water/fat separation problems. Unlabeled data are easier to obtain, but still it is difficult to ensure that training data do not lack test data characteristics. Accordingly, reconstruction directly from a test data without training is desirable, as in the reference T2*-IDEAL but without its dependence on initialization. This can be achieved using the proposed NTD.

The NTD method can overcome the initialization-dependence in traditional gradient descent based nonconvex optimization begs some intuitive explanation or interpretation, though rigorous explanation is currently not available. The cost function in DNN is minimized by adjusting network weights through backpropagation, which is achieved through iterative stochastic gradient descent (SGD). Perhaps SGD allows escaping local traps encountered in traditional gradient descent and the intensive computation in updating network weights facilitates some exhaustive search for a consistent minimum. The network weights updating on a single test data may converge as demonstrated in fidelity imposed network edit (17) and in deep image prior (16,19). Our data suggests that NTD can converge to a consistent minimum without initialization-dependence for the nonlinear nonconvex water/fat separation problem.

While some image details in STD, UTD, and NTD methods were different from reference images, quantitative measurements in the liver include ROI measurements of several voxels and regions which is less dependent on image details (20,21). The network architecture can be further optimized for this problem. For instance, we found changing activation function (Relu to Sigmoid) significantly improved field and R2* results in unsupervised training while not much change was observed in water and fat maps. Since field and R2* have a nonlinear relationship with respect to input complex signal while water and fat are linear, there is possibility of designing task-specific blocks to learn the linear and nonlinear mapping during training for further improvement and potentially decreasing the number of learned weights.

This study has a number of limitations. While the input data is complex, the network only accepts real values and two separate input channels were used instead. This potentially can change the noise properties and suboptimal network performance which can be addressed by including complex networks for this purpose (22). While the network in both supervised (STD) and the unsupervised (UTD) method were trained with specific acquisition parameters, it's possible to generalize by including more cases with different acquisitions parameters for training. Generation of reference images requires solving the T2*-IDEAL problem which can be challenging to calculate depending on acquisition protocol. The advantage of the proposed unsupervised methods is they only require complex input signal for training. While the NTD method requires a large number of epochs to converge which is computationally expensive, one could use a previously trained network (trained on data with the same imaging parameters) and update the weights for the new data set (23). Only one scanner make and model was used for this study and including additional models, manufacturers and field strengths will help to further generalize. There were limited number of cases used for training in STD and UTD methods in this study and including more cases will leverage network performance and reliability.

In summary, we demonstrated the feasibility of using unsupervised DNNs to solve the water/fat problem with very good agreement compared to reference images.

**REFERENCES:**

1. Reeder SB, Hu HH, Sirlin CB. Proton density fat-fraction: A standardized mr-based biomarker of tissue fat concentration. Journal of Magnetic Resonance Imaging 2012;36(5):1011-1014.

2. Labranche R, Gilbert G, Cerny M, Vu KN, Soulières D, Olivié D, Billiard JS, Yokoo T, Tang A. Liver Iron Quantification with MR Imaging: A Primer for Radiologists. Radiographics 2018;38(2):392-412.

3. Jafari R, Sheth S, Spincemaille P, Nguyen TD, Prince MR, Wen Y, Guo Y, Deh K, Liu Z, Margolis D, Brittenham GM, Kierans AS, Wang Y. Rapid automated liver quantitative susceptibility mapping. J Magn Reson Imaging 2019;50(3):725-732.

4. Sharma SD, Hernando D, Horng DE, Reeder SB. Quantitative susceptibility mapping in the abdomen as an imaging biomarker of hepatic iron overload. Magn Reson Med 2015;74(3):673-683.

5. Guo Y, Liu Z, Wen Y, Spincemaille P, Zhang H, Jafari R, Zhang S, Eskreis-Winkler S, Gillen KM, Yi P, Feng Q, Feng Y, Wang Y. Quantitative susceptibility mapping of the spine using in-phase echoes to initialize inhomogeneous field and R2* for the nonconvex optimization problem of fat-water separation. NMR in Biomedicine 2019;32(11):e4156.

6. Tsao J, Jiang Y. Hierarchical IDEAL: Fast, robust, and multiresolution separation of multiple chemical species from multiple echo times. Magnetic Resonance in Medicine 2013;70(1):155-159.

7. Zhong X, Nickel MD, Kannengiesser SAR, Dale BM, Kiefer B, Bashir MR. Liver fat quantification using a multi-step adaptive fitting approach with multi-echo GRE imaging. Magnetic Resonance in Medicine 2014;72(5):1353-1365.

8. Yu H, McKenzie CA, Shimakawa A, Vu AT, Brau AC, Beatty PJ, Pineda AR, Brittain JH, Reeder SB. Multiecho reconstruction for simultaneous water-fat decomposition and T2* estimation. J Magn Reson Imaging 2007;26(4):1153-1161.

9. Hernando D, Kellman P, Haldar JP, Liang ZP. Robust water/fat separation in the presence of large field inhomogeneities using a graph cut algorithm. Magn Reson Med 2010;63(1):79-90.

10. Dong J, Liu T, Chen F, Zhou D, Dimov A, Raj A, Cheng Q, Spincemaille P, Wang Y. Simultaneous phase unwrapping and removal of chemical shift (SPURS) using graph

**FIGURES:**

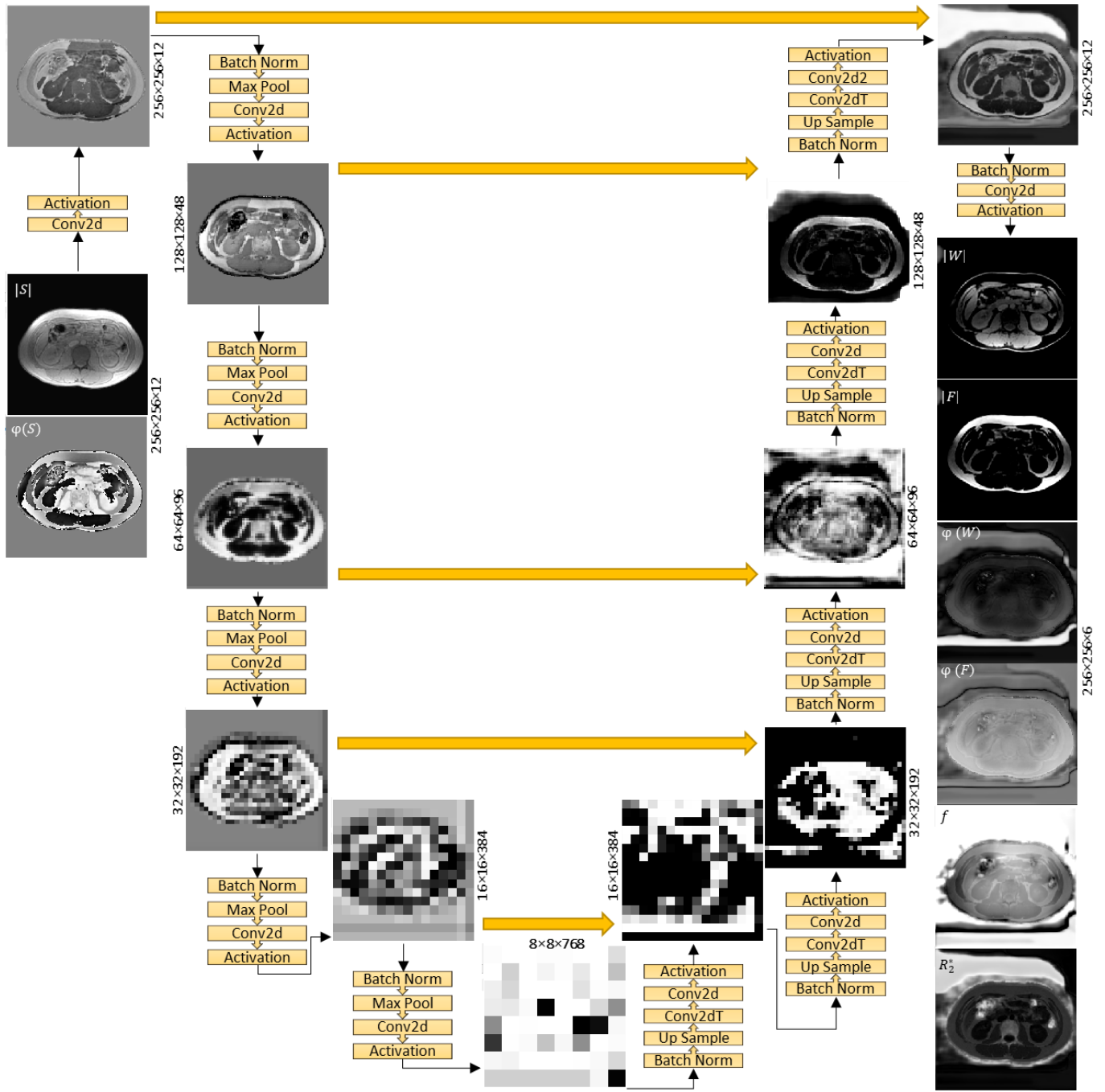

Figure 1. Network architecture, input complex signal magnitude and phase ($|S|, \varphi(S)$), intermediate layers output images, and network outputs including magnitude and phase of water($|W|, \varphi(W)$), magnitude and phase of fat ($|F|, \varphi(F)$), field ($f$), and $R_2^*$ maps are shown for a test dataset after training was completed.

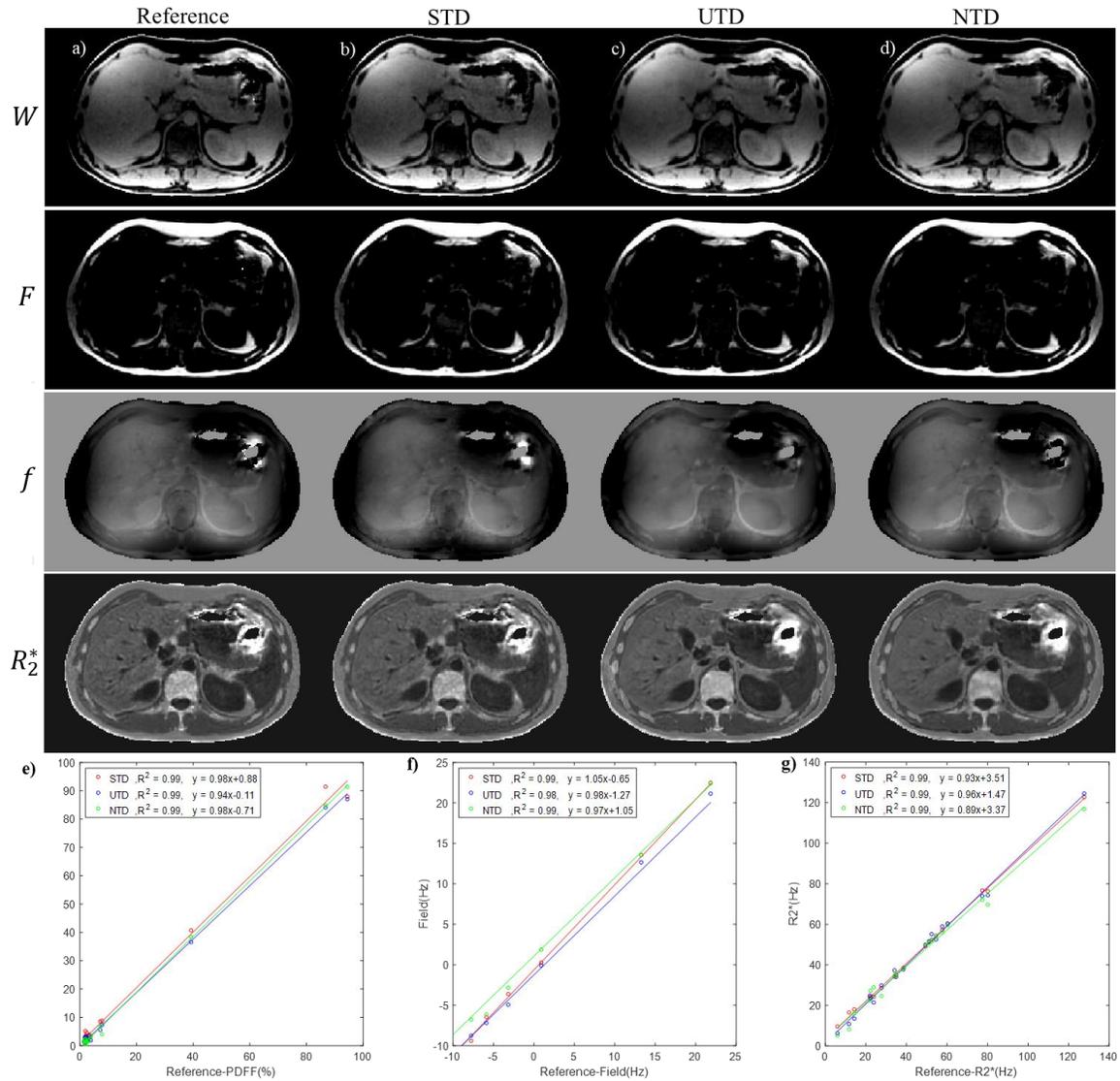

Figure 2. Water, fat, field and R2* reference images are shown (a) in a healthy volunteer. (b), (c), and (d) show corresponding results for supervised (STD), unsupervised (UTD), and NTD methods. ROI measurement correlation analysis shows excellent agreement between each DNN method and the reference T2*-IDEAL reconstruction for proton density fat fraction (e), field (f), and R2* (g).

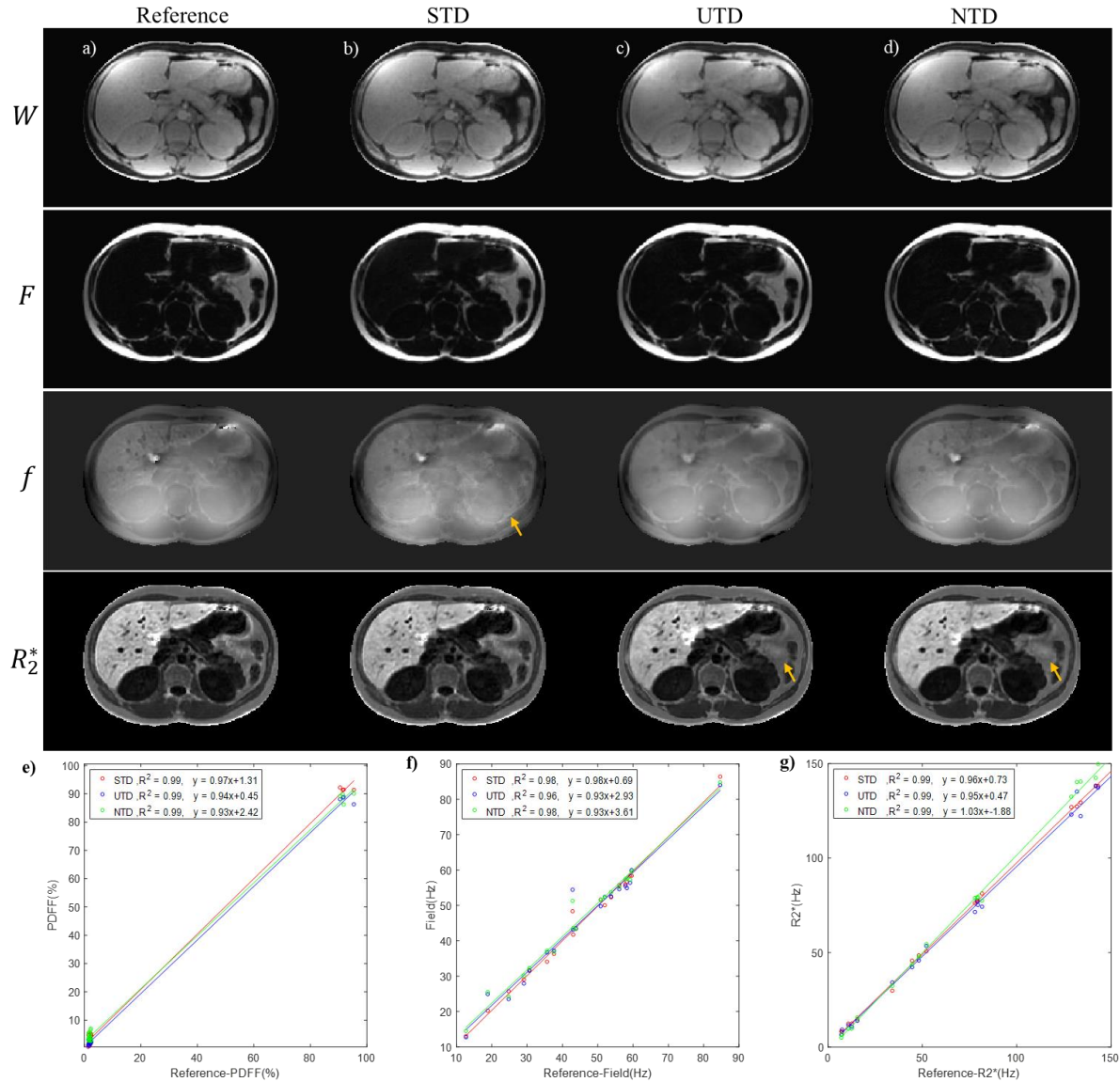

Figure 3. Water, fat, field and R2* reference images are shown (a) in a moderate iron-overload patient. (b), (c), and (d) show corresponding results for supervised (STD), unsupervised (UTD), and NTD methods. ROI measurement correlation analysis shows excellent agreement between each DNN method and the reference T2*-IDEAL reconstruction for proton density fat fraction (e), field (f), and R2* (g).

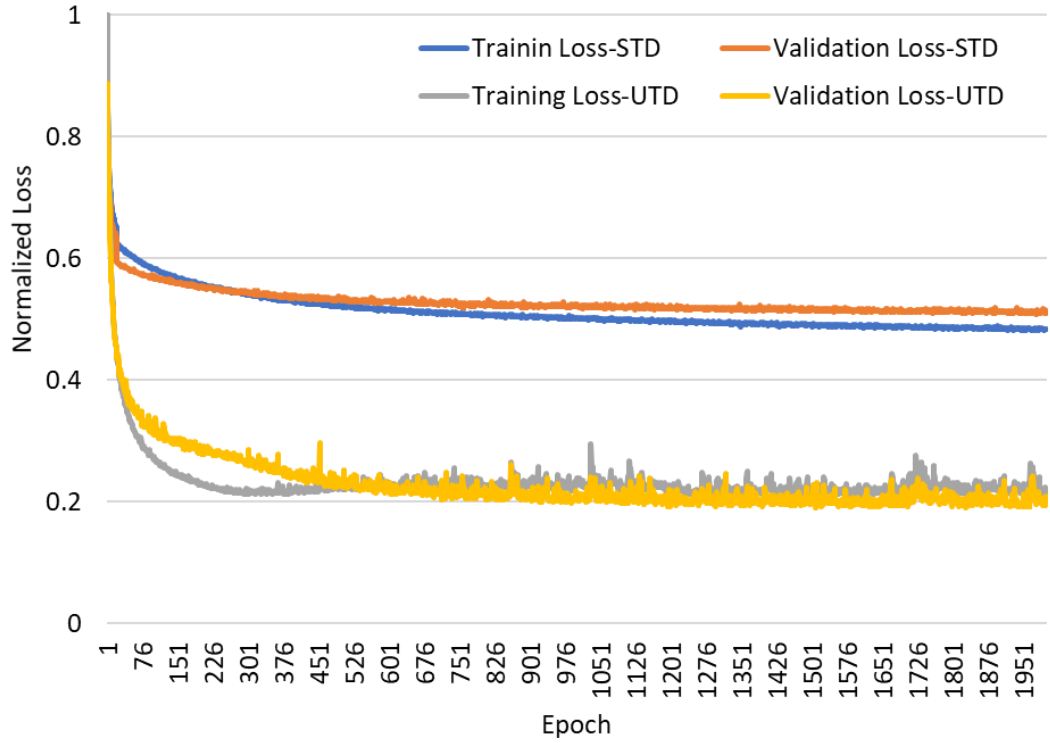

Figure 4 Training and validation loss results for supervised (STD) and unsupervised (UTD) training methods.

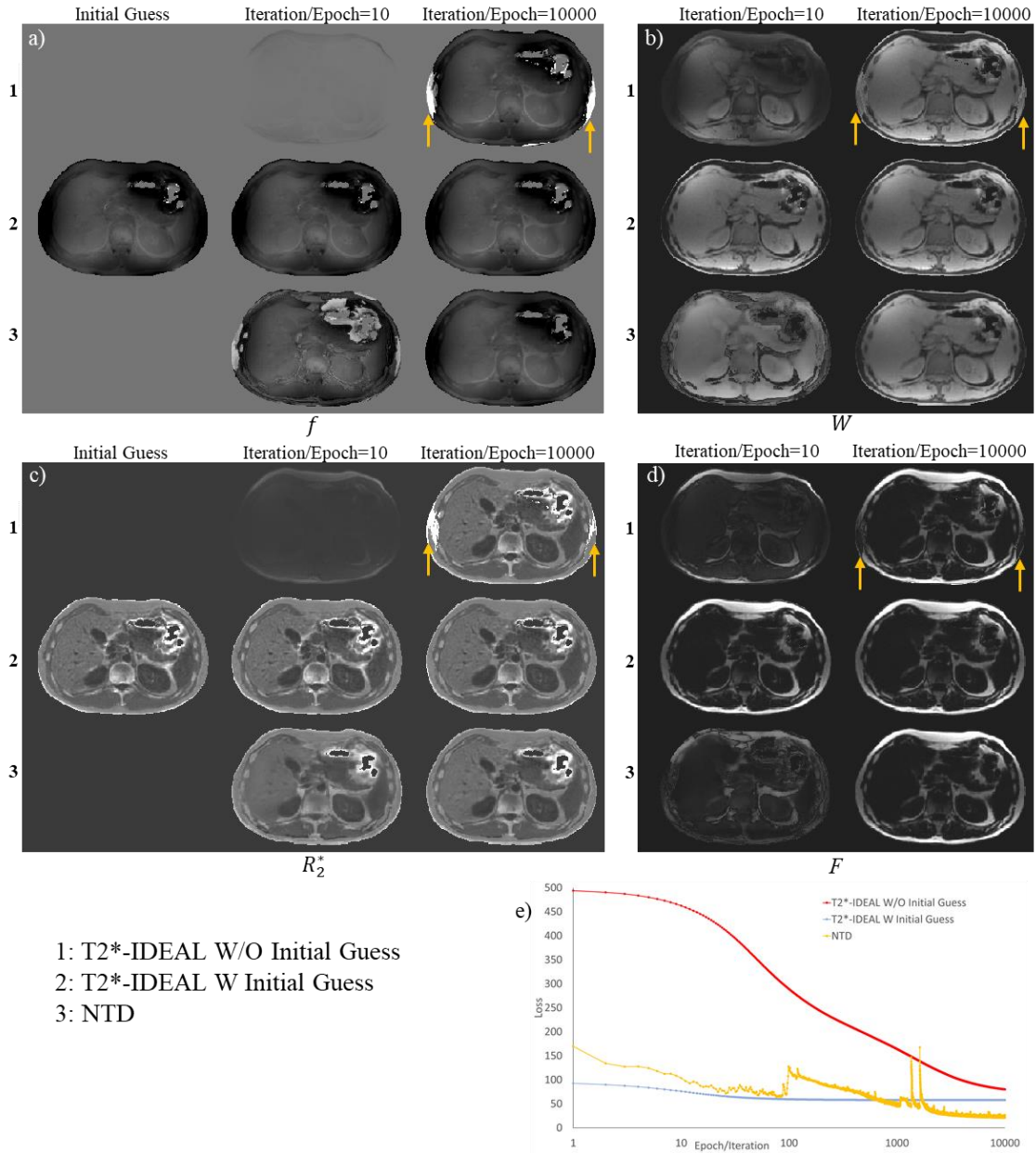

Figure 5. Water/fat separation results comparing current T2*-IDEAL without (1) and with (2) initialization with the proposed NTD method (3) training results for field (a), water (b), R2* (c), and fat (d). No initial guess was used for water and fat maps. Corresponding loss curves (e) are shown for T2*-IDEAL Vs. iteration and NTD Vs. epoch.